\documentclass[prl, aps, twocolumn, superscriptaddress, floatfix, nofootinbib, amsmath, amssymb, preprintnumbers]{revtex4-1}
\usepackage{graphicx}
\usepackage{dcolumn}
\usepackage{verbatim}
\usepackage{bm}
\usepackage{todonotes}
\usepackage{lmodern}
\usepackage{color}
\usepackage{colortbl}

\usepackage{tikz}
\usepackage{physics}
\usepackage{pgfplots}
\pgfplotsset{compat=newest,every axis plot/.append style={line width=1pt}}
\usepackage[colorlinks,linktocpage,linkcolor=cyan,citecolor=cyan]{hyperref}

\usepackage{tabularx}

\definecolor{lightgray}{gray}{0.9}
\definecolor{Amber}{rgb}{1.0, 0.75, 0.0}
\definecolor{blizzardblue}{rgb}{0.67, 0.9, 0.93}

\newcommand{\calR}{\mathcal{R}}
\def\fnl{f^{\text{local}}_{\text{NL}}}
\def\gnl{g^{\text{local}}_{\text{NL}}}

\newcommand{\rom}[1]{\uppercase\expandafter{\romannumeral #1\relax}}

\begin{document}

\preprint{YITP-21-143}

\title{One small step for an inflaton, one giant leap for inflation:\\
\it a novel non-Gaussian tail and primordial black holes}

\author{Yi-Fu Cai}
\email{yifucai@ustc.edu.cn}
\affiliation{Department of Astronomy, School of Physical Sciences, University of Science and Technology of China, Hefei, Anhui 230026, China}
\affiliation{CAS Key Laboratory for Researches in Galaxies and Cosmology, School of Astronomy and Space Science, University of Science and Technology of China, Hefei, Anhui 230026, China}

\author{Xiao-Han Ma}
\email{mxh171554@mail.ustc.edu.cn}
\affiliation{Department of Astronomy, School of Physical Sciences, University of Science and Technology of China, Hefei, Anhui 230026, China}
\affiliation{CAS Key Laboratory for Researches in Galaxies and Cosmology, School of Astronomy and Space Science, University of Science and Technology of China, Hefei, Anhui 230026, China}

\author{Misao Sasaki}
\email{misao.sasaki@ipmu.jp}
\affiliation{Kavli Institute for the Physics and Mathematics of the Universe (WPI), UTIAS, The University of Tokyo, Chiba 277-8583, Japan}
\affiliation{Center for Gravitational Physics, Yukawa Institute for Theoretical Physics, Kyoto University, Kyoto 606-8502, Japan}
\affiliation{Leung Center for Cosmology and Particle Astrophysics, National Taiwan University, Taipei 10617}

\author{Dong-Gang Wang}
\email{dgw36@cam.ac.uk}
\affiliation{Department of Applied Mathematics and Theoretical Physics, University of Cambridge, Wilberforce Road, Cambridge, CB3 0WA, UK}

\author{Zihan Zhou}
\email{zihanz@princeton.edu}
\affiliation{Department of Physics, Princeton University, Princeton, NJ 08544, USA}


\begin{abstract}
We report a novel prediction from single-field inflation that even a tiny step in the inflaton potential can change our perception of primordial non-Gaussianities of the curvature perturbation. Our analysis focuses on the tail of probability distribution generated by an upward step transition between two stages of slow-roll evolution. The nontrivial background dynamics with off-attractor behavior is identified. By using a non-perturbative $\delta N$ analysis, we explicitly show that a highly non-Gaussian tail can be generated by a tiny upward step, even when the conventional nonlinearity parameters $\fnl$, $\gnl$, etc. remain small. With this example, we demonstrate for the first time the sensitive dependence of non-perturbative effects on the tail of probability distribution. Our scenario has an inconceivable application to primordial black holes by either significantly boosting their abundance or completely forbidding their appearance.
\end{abstract}


\maketitle

{\it Introduction --}
Are primordial perturbations Gaussian? The latest cosmological observations suggest an affirmative answer by imposing tight constraints on primordial non-Gaussianities \cite{Planck:2018jri, Planck:2019kim}. Strictly speaking, however, the experiments of the cosmic microwave background (CMB) and large-scale structure (LSS) mainly probe fluctuations on large scales, while the non-Gaussian statistics on small scales is much less constrained. In particular, the statistics for large but rare perturbations on the tail of probability distribution remains unknown, even for those on the CMB or LSS scales. If deviations from the Gaussian distribution are present for these large perturbations, we expect rich and interesting phenomenology of {\it non-Gaussian tails}. Their implications have been actively discussed in the context of primordial black holes (PBHs) \cite{Franciolini:2018vbk, Biagetti:2018pjj, Atal:2018neu, Passaglia:2018ixg, Atal:2019cdz, Atal:2019erb, Ezquiaga:2019ftu, Figueroa:2020jkf, Pattison:2021oen, Taoso:2021uvl, Biagetti:2021eep, Davies:2021loj, Hooshangi:2021ubn}.

For the theoretical study of primordial perturbations, so far the efforts are mostly within the perturbative framework, where lower-point correlation functions, such as the bispectrum, are used as diagnostics of non-Gaussianities (see \cite{Chen:2010xka, Meerburg:2019qqi} for recent reviews). When perturbations are small, these correlation functions provide a successful description of non-Gaussian statistics. However, perturbative treatments break down for large and rare fluctuations. Accordingly, to properly understand  non-Gaussian tails, in principle one should go beyond perturbation theory, as recently discussed in \cite{Chen:2018uul, Chen:2018brw, Panagopoulos:2019ail, Panagopoulos:2020sxp, Celoria:2021vjw, Cohen:2021jbo}. Moreover, since non-perturbative effects may also play a significant role, we expect the perturbative and non-perturbative regimes are governed by different physics. Namely, rather counterintuitively, one may find a highly non-Gaussian tail even if the perturbative non-Gaussianities are small,which differs from the situation for the exponential tails extensively discussed in \cite{Biagetti:2018pjj, Ezquiaga:2019ftu, Figueroa:2020jkf, Pattison:2021oen, Biagetti:2021eep, Hooshangi:2021ubn}. In those cases, $\fnl$ and the non-Gaussian tails are normally correlated with each other, as the same physics governs both the perturbative and non-perturbative regimes.
%
%
Therefore, while slow-roll inflation models are normally seen as perturbative, it is interesting to explore non-perturbative processes that can completely alter the simplest scenario of inflation. In fact, for ultraviolet completions, inflationary landscapes are often expected to contain various deviations from slow-roll potentials \cite{Susskind:2003kw, Marsh:2015xka, Cai:2014vua, Palma:2017lww}. An intriguing question follows that how sensitive the non-Gaussian tails are to those non-perturbative effects.

In this Letter we take one simple but concrete example to show for the first time that, a small non-perturbative effect during inflation can drastically modify the tail of probability distribution. Our proposed mechanism is to introduce a tiny upward step along the slow-roll potential of the inflaton. When the inflaton climbs up this step, we observe transient off-attractor behavior similar to the non-attractor inflation \cite{Namjoo:2012aa, Chen:2013aj, Chen:2013eea, Cai:2016ngx, Cai:2017bxr}. A novel feature is that it gives rise to a highly non-perturbative phenomenon even if the step is extremely small. By using a non-perturbative $\delta N$ analysis, we observe that the nontrivial dynamics around the upward step can lead to a highly non-Gaussian tail while keeping the perturbative non-Gaussianities such as the bispectrum almost unchanged from the case without a step. As a direct application, we study PBHs generated from the upward step transition, and show explicitly that such a novel non-Gaussian tail can either significantly boost the PBH formation or completely shut it off. In this Letter we take $M_{\rm Pl}^2=1/8\pi G=1$.

\begin{figure}[htp!]
\centering
\includegraphics[width=0.9\columnwidth]{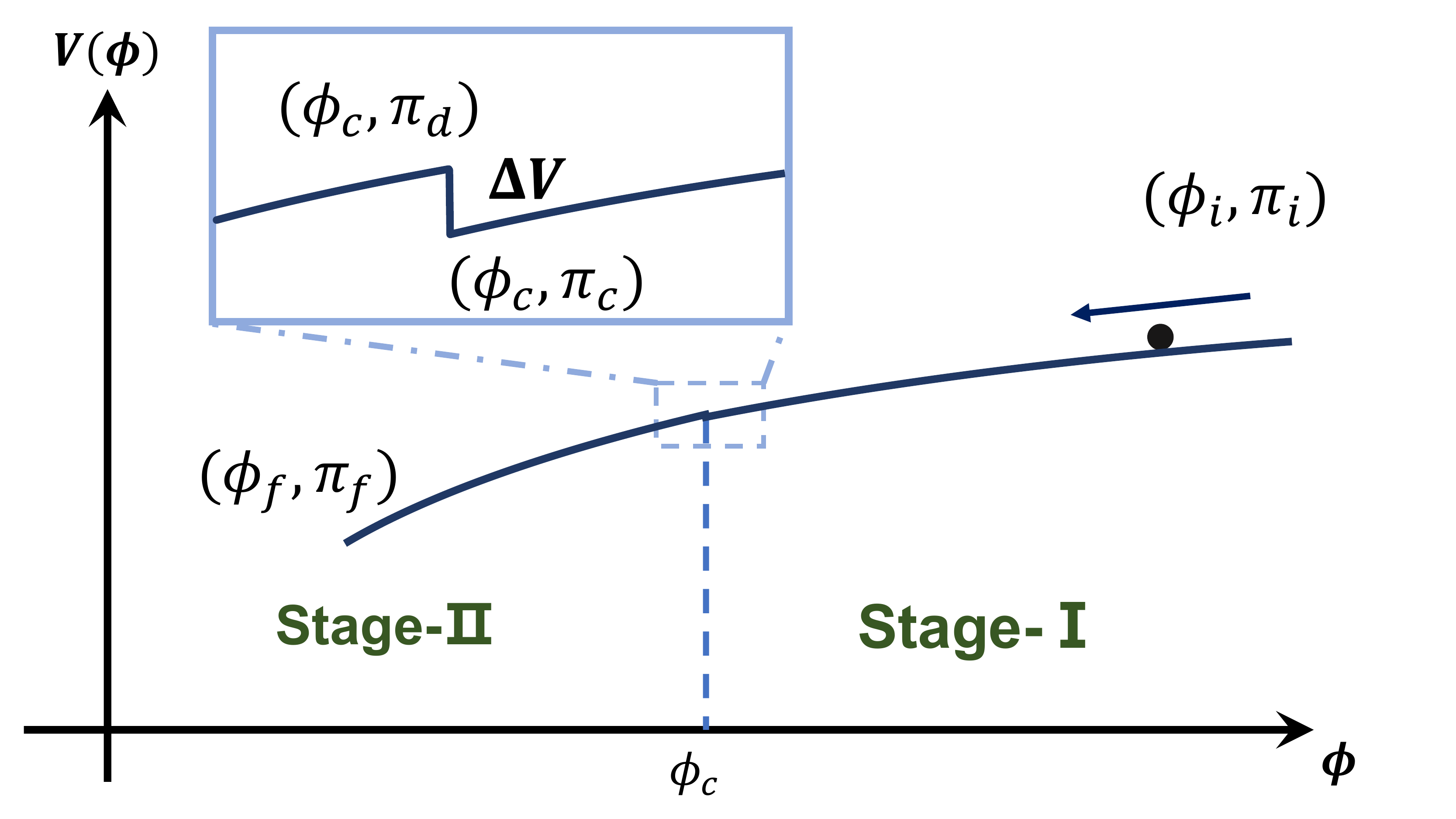}
\caption{A sketch of a potential for single-field inflation with a tiny upward step.}
\label{fig:SR_to_SR}
\end{figure}

{\it When the inflaton climbs up a step --}
We begin with a concrete example of single-field inflation where two stages of slow-roll evolution are connected by a tiny upward step in the potential, as shown in Fig.~\ref{fig:SR_to_SR}.
In this scenario, the inflaton starts rolling on the first slow-roll potential before reaching the step at $\phi_c$.
Afterwards, $\phi$ climbs up at a cost of partly losing its kinetic energy, and then undergoes a relaxation phase to initiate the second slow-roll stage.
As the transition process around the upward step is of the core interest, it is convenient to formulate the potential around the step by slow-roll parameters defined as $\epsilon_V \equiv \left(V'/V\right)^2/2$ and $\eta_V \equiv V''/V$.
The slow-roll parameters of the first and second stages may be
discretely different from each other.
For simplicity, however, we denote the slow-roll parameters by $\epsilon_V$ and $\eta_V$ for both stages unless confusion may arise.
The background dynamics of this scenario can be solved straightforwardly. Here we use the number of e-folds $n$ as the time variable via $dn = Hdt$.
At Stage-I, $\phi>\phi_c$, we have
\begin{align}
   \pi({n}) + \sqrt{2\epsilon_{V}}+
   \eta_{V}\left(\phi-\phi_{c}\right) \simeq0 \,;\quad \phi > \phi_c ~,
   \label{K-G equation of inflaton in phase1}
\end{align}
where $\pi \equiv d\phi/dn$ is the field velocity, which is completely determined by $\phi$, corresponding to an attractor trajectory in the phase-space diagram of $(\phi, \pi)$.

At the transition, $\phi=\phi_c$, denoting the field velocities before and after the step by $\pi_c$ and $\pi_d$, respectively, the energy conservation leads to the relation,
\begin{align}\label{pi_d}
    \pi_d = -\sqrt{\pi_c^2 - 6 ({\Delta V}/{V})} ~.
\end{align}
One sees that the inflaton can climb up the potential only if the step is small enough, $\Delta V/V< \pi_c^2/6$. For later convenience, we denote the ratio of these two velocities by
    $g\equiv {\pi_d}/{\pi_c}$.
It depicts how drastically the upward step affects the field velocity. The step is negligible for $g\rightarrow 1$, while it causes a significant drop in $\pi$ for $g\ll1$.

\begin{figure}[htp!]
\centering
\includegraphics[width=0.9\columnwidth]{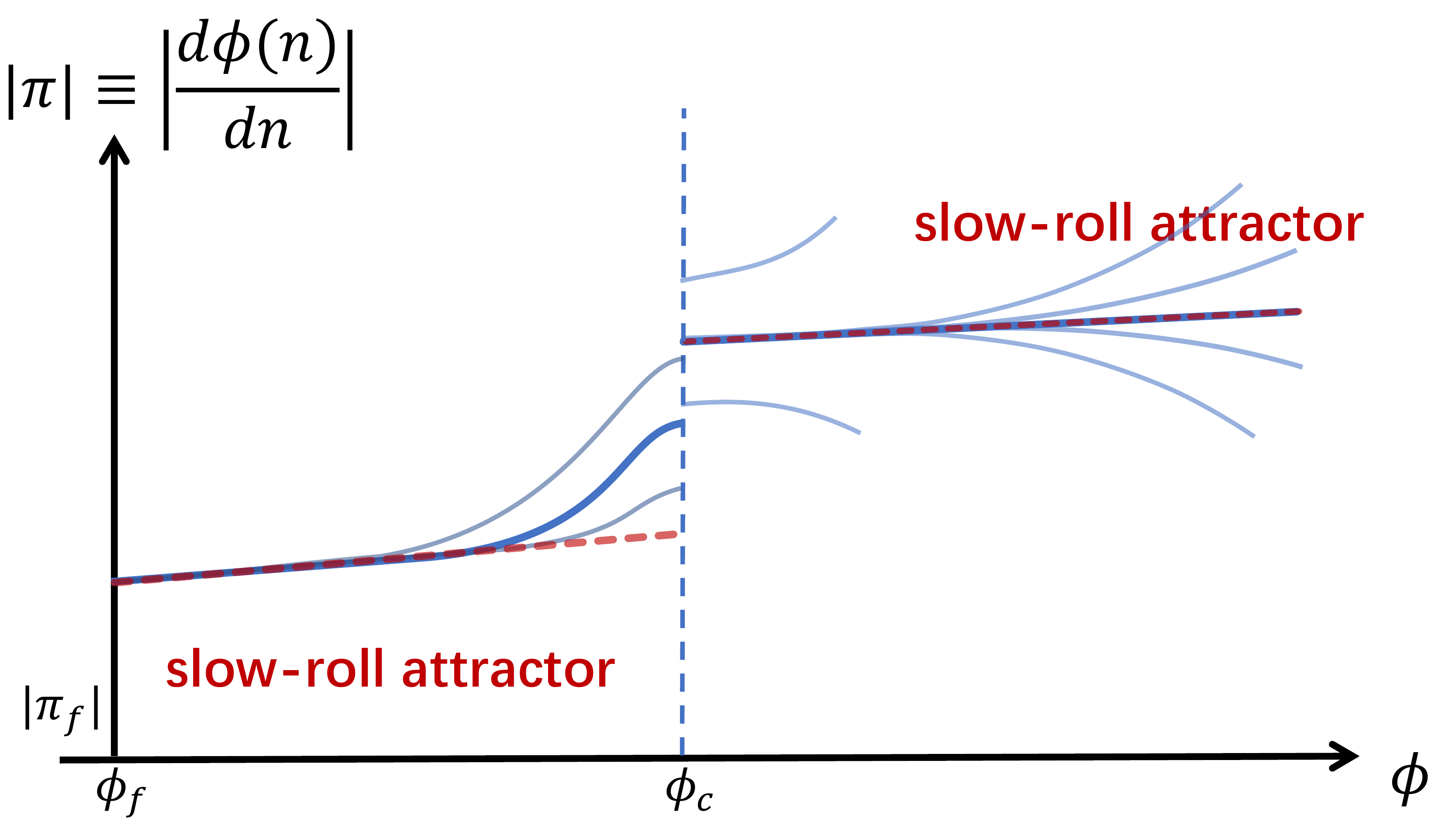}
\caption{Phase-space diagram $(\phi,\abs{\pi})$ near the step. The dashed red line is the slow-roll attractor. The thick blue curve is the base trajectory with slow-roll initial condition, while the light blue curves are the ones with off-attractor field velocities.}
\label{fig:phase_diagram}
\end{figure}

At the beginning of Stage-II, since the value of $\pi_d$ is different from the slow-roll attractor value, there appears a relaxation phase where the phase-space trajectory deviates from the attractor for a while.
Correspondingly, $\pi$ cannot be fully determined by $\phi$.
Within a few e-folds, however, the system approaches a slow-roll attractor phase again, which is given by
\begin{align}
    \phi_f-\phi_c \simeq\frac{\sqrt{2\epsilon_{V}}}{\eta_{V}}
\Bigg[ \Big(1+\frac{2\eta_{V}}{h}\Big) e^{-\eta_{V}(n_f-n_c)} -1 \Bigg] ~,
\label{phi_e approximate solution}
\end{align}
where $\phi_f=\phi(n_f)$ with $n_f$ being an epoch after the system has reached the attractor phase, and
\begin{align} \label{h}
 h \equiv \frac{6 \sqrt{2\epsilon_{V}}}{\pi_d}\simeq -6\frac{\pi_f}{\pi_d} ~.
\end{align}
Note that $\epsilon_V$ in $h$ is the one at the beginning of Stage-II.
The evolution of Stage-II is demonstrated by the phase-space trajectory in thick blue in Fig.~\ref{fig:phase_diagram}.

From the above background analysis, one might speculate that the cosmic expansion history is fully determined by a single dynamical degree of freedom $\phi(n)$, alike the standard slow-roll inflation. At first glance this seems reasonable, since during the whole evolution $\pi$ may be regarded as a function of $\phi$, which gives a fixed trajectory as shown by the thick blue curve in Fig.~\ref{fig:phase_diagram}. Even for the off-attractor phase at the beginning of Stage-II, $\pi$ is completely determined by the initial condition at $\phi=\phi_c$, which is determined by the attractor solution at $\phi>\phi_c$. Thus, one might conclude that the thick blue line in Fig.~\ref{fig:phase_diagram}, which we refer to {\it the base trajectory}, is the unique background trajectory for the upward step transition, and the corresponding prediction is not much different from the standard slow-roll inflation, apart from the transition effects on the vacuum fluctuations. However, the aforementioned argument has overlooked what may happen around the step. To see this clearly, we consider a perturbation of $\pi$ around the attractor solution at $\phi>\phi_c$. These {\it off-attractor trajectories} converge quickly to the slow-roll attractor, but there always exist off-attractor trajectories with $\pi_c$ not equal to the attractor value at $\phi=\phi_c$. Hence, $\pi_c$ is no longer fixed by the slow-roll dynamics in \eqref{K-G equation of inflaton in phase1}, and $\pi_d$ becomes a function of the off-attractor field velocity right before the step.
\begin{align}\label{non_attractor}
    \pi_c+3\phi_c = \pi_i + 3\phi_i.
\end{align}
Since $\pi_d$ provides the initial condition for the subsequent stages, the full evolution also becomes $\pi$-dependent, as demonstrated by light blue curves in Fig.~\ref{fig:phase_diagram}. Therefore, the base trajectory alone cannot provide a complete description of the whole dynamics. In the following, we testify the key role played by the off-attractor trajectories with the detailed perturbation analysis.

{\it A non-perturbative $\delta N$ analysis --}
We employ the $\delta N$ formalism \cite{Salopek:1990jq, Sasaki:1995aw, Starobinsky:1986fxa, Sasaki:1998ug, Lyth:2004gb, Lee:2005bb, Lyth:2005fi} to study the curvature perturbation to fully nonlinear order. This formalism is, by construction, applicable to any local type non-Gaussianities, which means
that it can be used to computed the non-Gaussianities from classical dynamics both in
perturbative and non-perturbative regimes. We focus on the modes which exit the Hubble radius before the transition. What we need is the total number of e-folds $N_{\text{tot}}$ from an epoch $n_i$ at Stage-I until $n=n_f$ when the inflaton has converged to the slow-roll attractor at Stage-II.
Adding contributions from the first and second stages, that are computed from the background solutions\eqref{non_attractor}  and  \eqref{phi_e approximate solution}, one gets
\begin{align}
    N_{\text{tot}} \simeq
    \frac{1}{3} \log\qty[ \frac{\pi_i}{\pi_c}]
    +\frac{\log( -2\eta_{V} \pi_d - 6\sqrt{2\epsilon_{V}})}{\eta_{V}}  + const ~.
\end{align}
As noted before, it is crucial that $\pi_d$ is not a fixed parameter but a function of $\phi_i$ and $\pi_i$. For a perturbed trajectory with initial condition, $(\phi_i + \delta\phi, \pi_i + \delta\pi)$, the relation \eqref{pi_d} gives
\begin{align}
    \pi_d+\delta\pi_d \simeq \pi_d
    \sqrt{1+\frac{6}{g}\frac{\delta\phi}{\pi_d} + 9\left(\frac{\delta\phi}{\pi_d}\right)^2} ~.
\label{pid-pert}
\end{align}
Varying $N_{\rm tot}$ with respect to $\phi_i$ and $\pi_i$ by using \eqref{pid-pert} yields the comoving curvature perturbation as
\begin{align}\label{nonpertR_Exact}
    \mathcal{R} \simeq
    -\frac{\delta \phi}{\pi_c} + \frac{2}{\abs{h}}
    \qty[ 1 - \sqrt{1+\frac{6}{g}\frac{\delta\phi}{\pi_d} + 9\left(\frac{\delta\phi}{\pi_d}\right)^2} ] ~,
\end{align}
where the step effects are mainly reflected in the square root. When the step vanishes ($g=1$), the square root disappears and we are left with $\calR =-(1+ 6/|h|)\delta\phi/\pi_d$.
But with the presence of a step ($g<1$), $\calR$ becomes a nonlinear function of the inflaton fluctuation $\delta \phi$, which
describes the fully non-perturbative regime as well as the perturbative regime.

In the perturbative regime where $|\calR|\ll1$, the Taylor expansion yields:
\begin{equation}
    \begin{aligned}
        \mathcal{R} \simeq&
        \qty[ \frac{6}{g^2 (h+2\eta_{V})} - 1 ] \frac{\delta\phi}{\pi_c} \\
        &+  \qty[ 9\frac{  \left(g^2-1\right) h+2 \eta_{V}\left(g^2-2\right)}{g^4 (h+2 \eta_{V})^2}+\frac{3}{2}] \qty(\frac{\delta\phi}{\pi_c})^2 + \cdots . \label{deltaNpert}
    \end{aligned}
\end{equation}
The leading terms consist of two contributions: one is the standard slow-roll result $-\delta\phi/\pi_c$, and the other due to the upward step. The latter dominates for the parameter regime, $g^2\abs{h}\ll 6$.
Assuming $g^2\ll1$, the local non-Gaussianity parameter $\fnl$ is given by
$\fnl\simeq 5|h|/12$. Since $h=-6\pi_f/\pi_d$ can be any negative value, it is possible to have large $\fnl$ in the upward step transition of single-field inflation.

Next, we consider the non-perturbative regime where $|\calR|$ is large.
Introducing the Gaussian part of curvature perturbation
$\mathcal{R}_{G} \equiv ({6}/{gh})({\delta\phi}/{\pi_d})$
and assuming $g^2\ll1$, $\calR$ can be approximated as
\begin{align}
 \mathcal{R} \simeq \frac{2}{\abs{h}} \big( 1 - \sqrt{1- |h| \mathcal{R}_{G} } \big) ~.
\label{nonpertR}
\end{align}
An intriguing fact is that $\cal R$ cannot be larger than $2/\abs{h}$.
Physically, this cutoff is due to the fact that the inflaton cannot climb up the step if the perturbation renders the absolute value of $\pi_c$ smaller than the critical value, $|\pi_c^{\rm crit}|=(\Delta V/6V)^{1/2}$.

{\it Non-Gaussian tails--}
The above result has a remarkable implication.
In the perturbative regime, one can always do Taylor expansions for small $\calR_{G}$. In the limit $|h| \ll 1$, the non-Gaussian coefficient at order $\calR_{G}^{1+n}$ is $O(|h|^n)$ for all $n\geq1$, which implies perturbative non-Gaussianities remain small in all orders.

\begin{figure}[htp!]
\centering
\includegraphics[width=\columnwidth]{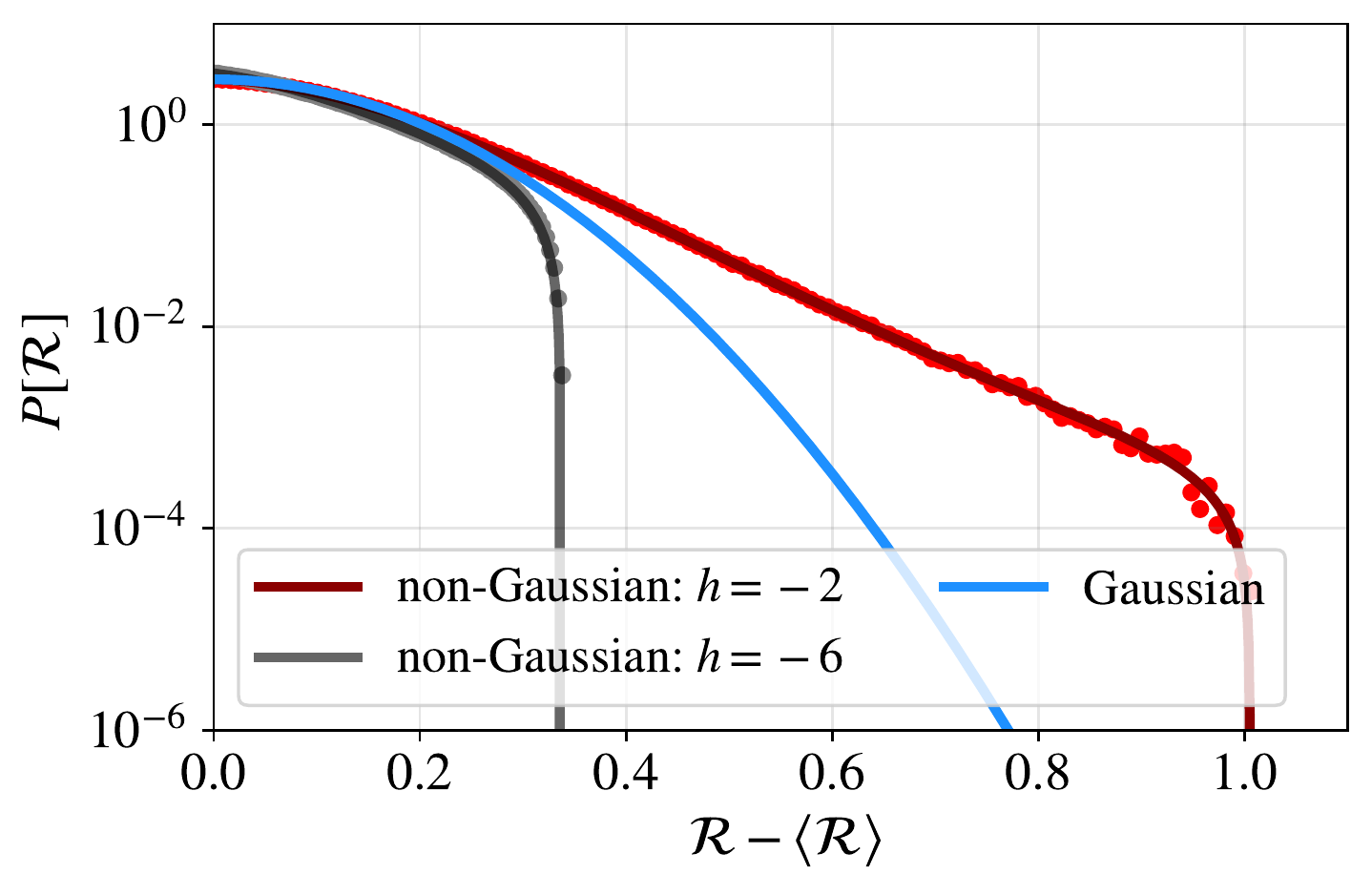}
\caption{Comparison among non-Gaussian and Gaussian $\calR$ with different $h$. The dots represent the PDF from Monte Carlo simulation $(g = 0.1)$ based on the exact expression of $\mathcal{R}$ in Eq.\eqref{nonpertR_Exact}.} 
\label{fig:pdf}
\end{figure}

However, once the non-perturbative regime is considered, $\calR$ automatically becomes highly non-Gaussian. To demonstrate the non-Gaussianity clearly, we examine the probability distribution function (PDF) of the curvature perturbation.
Using the fact that $P[\calR] d\calR = P[\calR_{{G}}] d\calR_{\rm G}$ where $P[\calR_{{G}}] = \exp[ -\calR_{{G}}^2 / (2\sigma_{\calR}^2) ] / (\sqrt{2\pi} \sigma_{\calR})$ with $\sigma_{\calR}^2 = \int d\log k ~ \mathcal{P}_{\calR_{\text{G}}}(k)$, the PDF of $\calR$ is given by
\begin{align}\label{pdf of calR}
 P[\mathcal{R}]
 = \frac{2-\abs{h}\mathcal{R}}{\Omega} \exp\Big[ -\frac{\mathcal{R}^2(4 -\abs{h} \mathcal{R})^2}{32 \sigma^{2}_{\mathcal{R}}} \Big] ~,
\end{align}
for $\mathcal{R} \leq {2}/{\abs{h}}$. Here $\Omega$ is a normalization coefficient, $\Omega \equiv \sqrt{2\pi\sigma^{2}_{\calR}} [1+ {\rm Erf}( {1}/({|h| \sqrt{2\sigma^2_{\calR}}}) ) ]$. The comparison between this PDF and the Gaussian one is shown in Fig.~\ref{fig:pdf}, where we have chosen  $\sigma_\calR^2=0.02$ for demonstration. As one can see, in the perturbative regime the PDF behaves like Gaussian. When $|\calR|$ is large, the non-Gaussianity becomes prominent, with
a cutoff at $\calR = 2/|h|$.

The parameter $h$ plays an important role in the non-Gaussian PDF. For $|h|\gg1$, one gets a large local non-Gaussianity around $\calR\simeq 0$, as shown by the black curve in Fig.~\ref{fig:pdf}, and the tail of the distribution is highly suppressed as $\calR \rightarrow 2/|h|$. For $|h|\lesssim 1$, the non-Gaussianity in the perturbative regime is small. However, the non-Gaussian tail can be substantially enhanced for moderately large values of $\calR$, while the cutoff appears rather abruptly at $\calR=2/|h|$, as shown by the red curve in Fig.~\ref{fig:pdf}. The lesson here is that even if the non-Gaussianity is perturbatively small, the behavior for large and rare perturbations at the tail of the distribution can be highly non-perturbative, leading to a completely different picture from what one would expect in perturbative analysis.

{\it Formation of PBHs --} An interesting application of the non-Gaussian tail is the PBH formation. As is well-known, PBHs are formed when a density perturbation exceeds a critical value when the scale re-enters the Hubble radius. Thus their abundance is highly sensitive to the PDF tail for the density perturbation. To estimate the abundance of PBHs, one needs to first compute the power spectrum of $\calR$ in the upward step transition. Such enhanced power spectra has been obtained in \cite{Inomata:2021uqj, Inomata:2021tpx} for generating PBHs from potential steps. Here we particularly focus the
effect of the non-Gaussian tail on the PBH formation.

\begin{figure}[htp!]
\centering
\includegraphics[width=1\columnwidth]{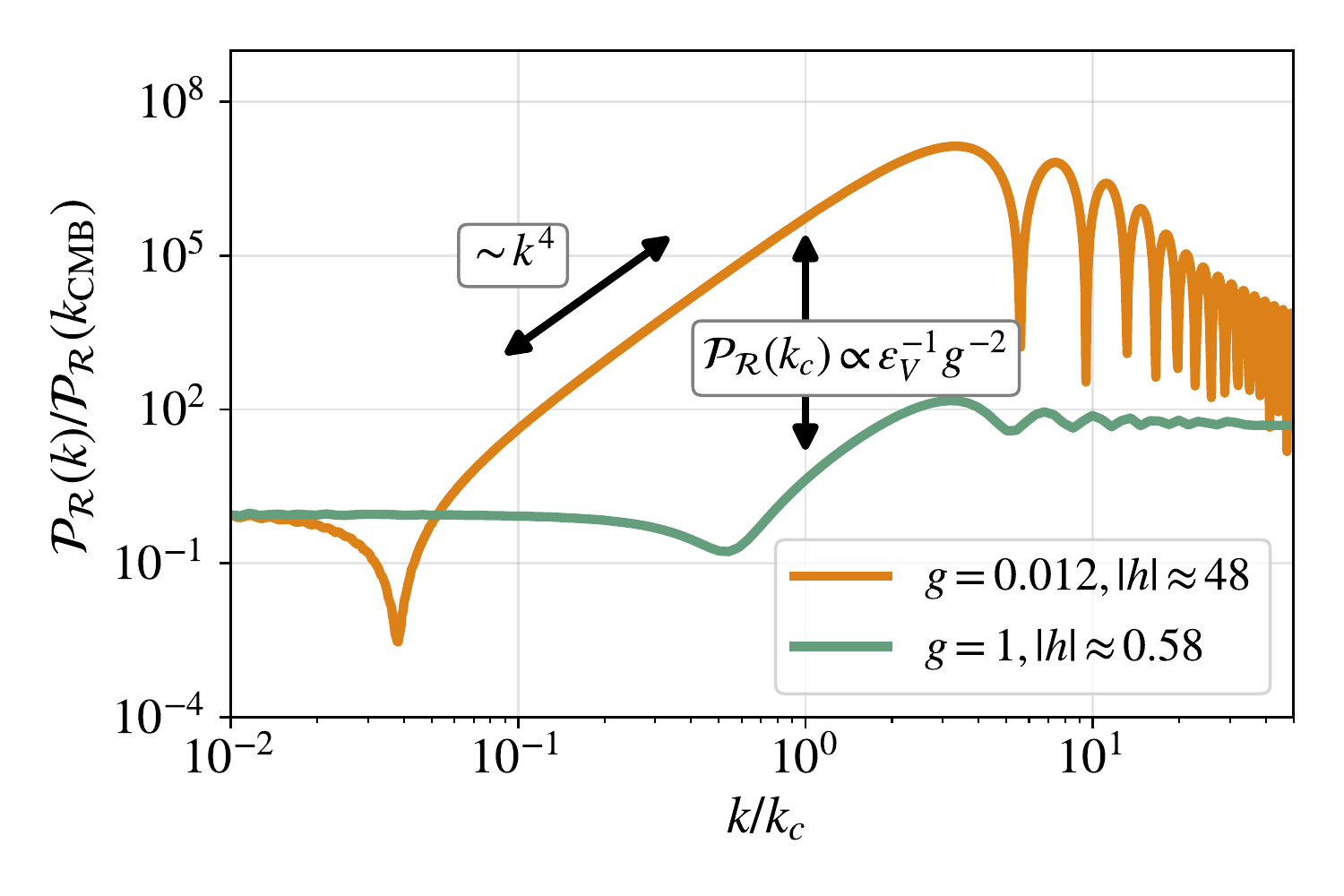}
\caption{
The power spectra from transitions with (orange) and without (green) an upward step. Except for the height of the step, the potential  parameters are the same for both cases.}
\label{fig:powerspectrum}
\end{figure}

As one may anticipate from the PDF in \eqref{pdf of calR}, or from Fig.~\ref{fig:pdf}, the curvature perturbation spectrum can be well approximated by that of the Gaussian part as long as $|h|\sigma_\calR$ is small. Then it can be computed straightforwardly by solving the linear perturbation equation for $\calR$ in Fourier space. During inflation, the initial condition for the mode $\calR_k$ is identified well as the adiabatic vacuum positive frequency function, $\mathcal{R}_{k}(\tau) = e^{-i k \tau} / (2a \sqrt{\epsilon k} )$, where $a=(-H\tau)^{-1}$. The upward step at $\phi_c$ induces a sudden change of in the time derivative of the slow-roll parameter $\epsilon$,
which causes a mixing of positive and negative frequencies. Denoting the slow-roll value of $\eta \equiv \dot\epsilon/(\epsilon H)$ right before the step at $\phi_c$ by $\eta_c$,
\footnote{To avoid possible misunderstanding, we note $\eta_c$ is the Hubble slow-roll parameter, not equal to $\eta_V$ defined by the potential.}
\begin{align}
\begin{aligned}
    \mathcal{P}_\mathcal{R}(k) & \simeq
    \mathcal{P}_\mathcal{R}^{\rm sr}(k)\frac{   \left(\eta_c+2 k^2 \tau_c^2\right)^2 +\eta_c^2 k^2 \tau_c^2}{ 4g^2  k^6 \tau_c^6} \\
    &\times\big[\sin (k \tau_c)-k \tau_c \cos (k\tau_c)\big]^2 ~,
\end{aligned}
\end{align}
where $\mathcal{P}_\mathcal{R}^{\rm sr}(k) = H^2 / ( 8\pi^2\epsilon_{V} )$ and $\tau_c$ is the conformal time at $\phi_c$. The $k$-dependence of the power spectrum can be analyzed in three different scales. The first one corresponds to the long wavelength modes with $k^2\tau_c^2\ll |\eta_c|\ll 1$, which gives a nearly scale-invariant power spectrum $\mathcal{P}_\mathcal{R}(k) \simeq \mathcal{P}_\mathcal{R}^{\rm sr}(k) ({\eta_c^2}/{36g^2})$.
For the short wavelength modes $-k\tau_c\gg1$, we find $\mathcal{P}_\mathcal{R}(k) \simeq \mathcal{P}_\mathcal{R}^{\rm sr}(k) \cos^2(k\tau_c)/g^2$. Here the constant oscillation is caused by the instantaneous transition and is expected to damp out as soon as we consider a nonzero duration of the transition. There is also an intermediate regime with $|\eta_c|< k^2\tau_c^2< 1$, and $\mathcal{P}_\mathcal{R}(k) \simeq \mathcal{P}_\mathcal{R}^{\rm sr}(k){k^4\tau_c^4}/{(9g^2)}$ which demonstrates the $k^4$ growth of the power spectrum in canonical single-field inflation \cite{Byrnes:2018txb,Carrilho:2019oqg,Ozsoy:2019lyy}.
A full numerical computation of the power spectrum is given in Fig.~\ref{fig:powerspectrum}, which confirms the above arguments. Note also that the power spectrum is peaked around $k_c = -1/\tau_c$, with the amplitude
$\mathcal{P}_{\mathcal{R}}(k_{c}) \simeq   \mathcal{P}_\mathcal{R}^{\rm sr}(k) /g^2$. Compared with the spectrum at long wavelength modes, the amplitude at the peak is enhanced by the factor $36/\eta_c^2$.

With the enhanced power spectrum and non-Gaussian tail, one can estimate the mass fraction $\beta(M)$ of PBHs at the formation time. Typically, the threshold for the PBH formation is given by a critical density perturbation on the comoving slicing, $\delta\rho^{\rm com} / \rho \equiv \Delta > \Delta_{cr}$, where $\Delta_{cr} \sim 0.5$. Although the relation between the curvature perturbation $\calR$ and density perturbation $\Delta$ is non-local in the sense that $\Delta$ is proportional to the spatial Laplacian of $\calR$ (see \cite{Young:2022phe} and \cite{DeLuca:2022rfz} for more detailed discussions) , assuming the spectrum is peaked, which is in fact the case in the current scenario, one can approximately assume $\Delta \sim k_c^2 / (a^2 H^2) \calR$, where $k_c$ is the peak wavenumber. Thus, $\Delta\sim\calR$ at the Hubble-crossing $k=aH$. For simplicity, we adopt this approximation and assume that $\beta(M)$ can be estimated by introducing the corresponding threshold $\calR_c$ of $\mathcal{O}(1)$ as $ \beta_{\text{PBH}} = \int^{2/|h|}_{\mathcal{R}_c} d \mathcal{R} P[\mathcal{R}]$. Using \eqref{pdf of calR}, we derive
\begin{align}
\begin{aligned}
    \beta^{\text{NG}}_{\text{PBH}}
    & = \frac{\sqrt{2\pi\sigma^{2}_{\calR}}}{\Omega} \Big[ \text{Erf}\big( \frac{1}{\abs{h}\sqrt{2\sigma^{2}_{\calR}}} \big) \\
    & - \text{Erf}\big( \frac{\calR_c(4-\abs{h}\calR_c )}{4\sqrt{2\sigma^{2}_{\calR}}} \big) \Big]
    \Theta\big( \frac{2}{\abs{h}}-\calR_c \big) ~.
    \label{NG mass fraction in calR}
\end{aligned}
\end{align}
For comparison, the mass fraction from the Gaussian PDF is $\beta^{\text{G}}_{\text{PBH}} = \frac{1}{2} \left[1-\text{Erf}( {\calR_c} / {\sqrt{2 \sigma^{2}_{\calR}}} ) \right]$. It is helpful to look at the ratio of the non-Gaussian and Gaussian results. To illustrate, we set $\calR_c=0.7$, and then, $\beta^{\text{NG}}_{\text{PBH}} / \beta^{\text{G}}_{\text{PBH}} \sim 10^3$ for $h=-2$; however, $\beta^{\rm NG} = 0$ if $|h|>2.8572$. Thus, we find that the non-Gaussian tail can either easily enhance the PBH mass fraction by several orders of magnitude, or make it absolutely impossible to form PBHs.

{\it Concluding remarks --}
In this Letter, we demonstrated for the first time that even a tiny upward step along the potential yields a significant effect on the tail of probability distribution for the curvature perturbation. Our analysis differs substantially from those on featured potential models in the literature, such as step features in \cite{Chen:2006xjb,Adshead:2011jq,Cai:2015xla,Inomata:2021uqj,Inomata:2021tpx}, which were based on the single-clock assumption, hence neglected the effects of off-attractor trajectories. The origin of the non-Gaussian statistics in our scenario is in analogue with the one in non-attractor inflation \cite{Namjoo:2012aa, Chen:2013aj, Chen:2013eea,Cai:2016ngx,Cai:2017bxr}, in that the position and velocity of the background field are two independent degrees of freedom, which consequently leads to non-Gaussianities distinct from single-clock models. One key novelty in our scenario is that a highly non-Gaussian tail arises non-perturbatively even if the perturbative non-Gaussianities such as $\fnl$ are small, which essentially differs from the exponential tails discussed in previous works. While in this Letter we focused on the simplest example to demonstrate key features, the detailed analyses of more realistic models are left in a follow-up study \cite{long}. There, by taking into account more model parameters and using detailed numerical computations, we further confirm the robustness of main results reported in this Letter.

Our work shed new lights on several research lines of the very early Universe. Theoretically, the novel non-Gaussian tail identified in the PDF of our mechanism has opened up a brand new window for theory of cosmological perturbations. With the specific example studied in this Letter, we certified that the probability distribution tail is highly sensitive to non-perturbative effects during inflation. Observationally, to deepen our knowledge of the Universe, it is essential to develop new approaches to nontrivial statistics beyond the perturbative regime to match with rapidly developing astronomical observations that could discover primordial non-Gaussianities in the near future. Moreover, there are plentiful non-perturbative phenomenologies to explore. For example, various resonances during inflation have been studied extensively in \cite{Cai:2018tuh, Cai:2019jah, Chen:2019zza, Chen:2020uhe, Zhou:2020kkf, Cai:2021wzd, Cai:2021yvq, Peng:2021zon}; one may expect nontrivial signatures from stochastic effects \cite{Ezquiaga:2019ftu, Figueroa:2020jkf, Pattison:2021oen} and multifield dynamics \cite{Pi:2021dft, Achucarro:2021pdh} during inflation. It is interesting to investigate rich behaviors of their non-Gaussian tails and other cases in follow-up studies.

We end by commenting that, our proposed scenario offers new perspectives towards the non-perturbative understanding of the Universe. Although the specific example is merely one small step for an inflaton, a giant leap is accomplished for the perception of inflation by probing the rarely explored zone of non-perturbative effects including a novel non-Gaussian tail and primordial black holes as well as other cosmological applications.

\

\begin{acknowledgments}
{\it Acknowledgments.--}
We are grateful to Chao Chen, Xingang Chen, Bichu Li, Chunshan Lin, Mohammad Hossein Namjoo, Bo Wang and Yi Wang for discussions.
YFC and XHM are supported in part by the National Key R\&D Program of China (2021YFC2203100), by the NSFC (11961131007, 11653002), by the Fundamental Research Funds for Central Universities, by the CSC Innovation Talent Funds, by the CAS project for young scientists in basic research (YSBR-006), by the USTC Fellowship for International Cooperation, and by the USTC Research Funds of the Double First-Class Initiative.
DGW is supported by the Rubicon Postdoctoral Fellowship awarded by the Netherlands Organisation for Scientific Research (NWO).
MS is supported in part by JSPS KAKENHI grants (19H01895, 20H04727, 20H05853), and by the World Premier International Research Center Initiative (WPI Initiative), MEXT, Japan.
ZZ is supported in part by the scholarship at Princeton.
We acknowledge the use of computing facilities of Kavli IPMU, as well as the clusters {\it LINDA} and {\it JUDY} of the particle cosmology group at USTC.
\end{acknowledgments}

\bibliography{references}

\end{document}